\documentclass[twocolumn,aps,prx,floatfix,superscriptaddress]{revtex4-1}

\usepackage{graphicx}
\usepackage{dcolumn}
\usepackage{epsfig}
\usepackage{amssymb}
\usepackage{amsmath}
\usepackage{natbib}
\usepackage{array}
\usepackage{bbold}
\usepackage{color}

\begin{document}
\title{Fine structure of $\mathrm{K}$-excitons in multilayers of transition metal dichalcogenides}
\author{A. O. Slobodeniuk}
\email{artur.slobodeniuk@lncmi.cnrs.fr}
\affiliation{Laboratoire National des Champs Magn\'etiques Intenses, CNRS-UGA-UPS-INSA-EMFL, 25 avenue des Martyrs, 38042 Grenoble, France}
\author{\L{}. Bala}
\affiliation{Laboratoire National des Champs Magn\'etiques Intenses, CNRS-UGA-UPS-INSA-EMFL, 25 avenue des Martyrs, 38042 Grenoble, France}
\affiliation{Institute of Experimental Physics, Faculty of Physics, University of Warsaw, ul. Pasteura 5, 02-093 Warszawa, Poland}
\author{M. Koperski}
\affiliation{Laboratoire National des Champs Magn\'etiques Intenses, CNRS-UGA-UPS-INSA-EMFL, 25 avenue des Martyrs, 38042 Grenoble, France}
\affiliation{School of Physics and Astronomy, University of Manchester, Oxford Road, Manchester, M13 9PL, UK}
\affiliation{National Graphene Institute, University of Manchester, Oxford Road, Manchester, M13 9PL, UK}
\author{M. R. Molas}
\affiliation{Laboratoire National des Champs Magn\'etiques Intenses, CNRS-UGA-UPS-INSA-EMFL, 25 avenue des Martyrs, 38042 Grenoble, France}
\affiliation{Institute of Experimental Physics, Faculty of Physics, University of Warsaw, ul. Pasteura 5, 02-093 Warszawa, Poland}
\author{P. Kossacki}
\affiliation{Institute of Experimental Physics, Faculty of Physics, University of Warsaw, ul. Pasteura 5, 02-093 Warszawa, Poland}
\author{K.~Nogajewski}
\affiliation{Laboratoire National des Champs Magn\'etiques Intenses, CNRS-UGA-UPS-INSA-EMFL, 25 avenue des Martyrs, 38042 Grenoble, France}
\affiliation{Institute of Experimental Physics, Faculty of Physics, University of Warsaw, ul. Pasteura 5, 02-093 Warszawa, Poland}
\author{M.~Bartos}
\affiliation{Laboratoire National des Champs Magn\'etiques Intenses, CNRS-UGA-UPS-INSA-EMFL, 25 avenue des Martyrs, 38042 Grenoble, France}
\author{K. Watanabe}
\affiliation{National Institute for Materials Science, 1-1 Namiki, Tsukuba 305-0044, Japan}
\author{T. Taniguchi}
\affiliation{National Institute for Materials Science, 1-1 Namiki, Tsukuba 305-0044, Japan}
\author{C. Faugeras}
\affiliation{Laboratoire National des Champs Magn\'etiques Intenses, CNRS-UGA-UPS-INSA-EMFL, 25 avenue des Martyrs, 38042 Grenoble, France}
\author{M. Potemski}
\email{marek.potemski@lncmi.cnrs.fr}
\affiliation{Laboratoire National des Champs Magn\'etiques Intenses, CNRS-UGA-UPS-INSA-EMFL, 25 avenue des Martyrs, 38042 Grenoble, France}
\affiliation{Institute of Experimental Physics, Faculty of Physics, University of Warsaw, ul. Pasteura 5, 02-093 Warszawa, Poland}

\begin{abstract}

Reflectance and magneto-reflectance experiments together with theoretical modelling based on the $\mathbf{k\cdot p}$ approach have been employed to study the evolution of direct bandgap excitons in MoS$_2$ layers with a thickness ranging from mono- to trilayer. The extra excitonic resonances observed in MoS$_2$ multilayers emerge as a result of the hybridization of Bloch states of each sub-layer due to the interlayer coupling. The properties of such excitons in bi- and trilayers are classified by the symmetry of corresponding crystals. The inter- and intralayer character of the reported excitonic resonances is fingerprinted with the magneto-optical measurements: the excitonic $g$-factors of opposite sign and of different amplitude are revealed for these two types of resonances. The parameters describing the strength of the spin-orbit interaction are estimated for bi- and trilayer MoS$_2$.

\end{abstract}

\maketitle

\section{Introduction}

Scientific curiosity to uncover the properties of new materials and to demonstrate their possible novel functionalities drive
the research efforts focused on atomically-thin matter, and, in particular, on thin layers of semiconducting transition metal
dichalcogenides (S-TMD)\cite{novoselov2005,wang2012,koperski,Urbaszek2018}. Intense works have been devoted to
studies of S-TMD monolayers which appeared to be the
efficient light emitters, the two-dimensional semiconductors with a direct bandgap positioned at the $\mathrm{K}^\pm$ points
of their 1-st hexagonal Brillouin zone (BZ) \cite{mak2010,binder2017,Palacio2016}. New and rich possibilities of tuning the band structure, the strength of
Coulomb interaction, and thus the optical properties, are opened when stacking the S-TMD monolayers into a form of
multilayers and/or hetero-layers \cite{ugeda2014,ye,chernikov2015,stierhBN,Withers2015,
Withers2015NN,Genevie2016,schwarz2016,Shinde2018}. The properties of the archetypes of S-TMD stacks which are the thermodynamically
stable 2H-stacked multilayers are to be well understood first.

In 2H stacks of $N$ monolayers ($N$ML), the electronic bands are known to be effectively modified, with $N$, in the range outside the $\mathrm{K}^\pm$ points of the BZ \cite{Tijerina2018,sun2016,bradley2015,fang2015,cheiw2012,padilha2014,debbichi2014}. This, in particular, implies the indirect bandgap in $N$MLs when
$N>1$, what strongly affects the emission spectra of these multilayers \cite{splendiani2010,arora2015,aroramose2,molasNanoscale,zeng2013,zhang2014}.
Instead, more subtle effects of the
hybridization of electronic states around the direct bandgap which appears at $\mathrm{K}^\pm$ points of the BZ in any $N$ML
are relevant for the absorption-type processes \cite{gong_kp,kormanyos2018,komsa2013}. Understanding the absorption response of S-TMD multilayers
might be of special importance for their potential applications in photo-sensing or photo-voltaic devices \cite{bernardi2013,desai2014,dhakal2014,zhao2013}.

In this paper we present the theoretical outline (based on $\mathbf{k\cdot p}$ approach \cite{Tijerina2018,gong_kp,molasNanoscale,Arora2018}) and the experimental data
(results of reflectance and magneto-reflectance measurements), which, in a consistent manner, unveil the nature of direct
bandgap ($\mathrm{K}^\pm$ points) excitonic transitions in 2H-stacked S-TMDs multilayers. The geometry of 2H stacking,
wherein each subsequent layer is $180^\circ$ rotated around previous one, induces interaction and hybridization of the
Bloch states in $\mathrm{K}^+/\mathrm{K}^-/\mathrm{K}^+\ldots(\mathrm{K}^-/\mathrm{K}^+/\mathrm{K}^-\ldots)$  points of
subsequent monolayers, which form the $\mathrm{K}^+(\mathrm{K}^-)$ valleys of the multilayer. Such non-trivial interlayer
coupling delocalizes the electron states in the out-of-plane direction, which leads to the formation of the new type of
excitons, associated with $\mathrm{K}^\pm$ valleys of multilayers as well as bulk crystals \cite{molasNanoscale,horng2018,arora2017}.
We classify such excitons, associate them with the symmetry of the crystal and describe their properties in terms of the
simple theoretical model. Notably, the intra- or interlayer character of the excitonic transitions is fingerprinted with,
correspondingly, positive or negative sign of the $g$-factor associated to these transitions.

Due to the different symmetries of the conduction and valence band orbitals, the hybridization of $\mathrm{K}^\pm$-electronic
states in $N$ML occurs predominantly in the valence band and is more effective when the spin-orbit splitting $\Delta_v$
in the valence band is small. MoS$_2$ crystals, with the smallest $\Delta_v$ among all other S-TMDs, have been therefore chosen
for the investigations. In the experiments, we have largely profited of the significantly improved optical quality of MoS$_2$
layers when they are encapsulated in between the hexagonal boron nitride (hBN)\cite{cadizMoS2,wierzbowski2017,wang2017,vaclavkova,
stierhBN,chen2018}. We consider the bi- and trilayer systems
as the simplest multilayer representatives with different spatial symmetry. Comparing the experimental data with the theoretical
expectation on a more quantitative level, we discuss the characteristic parameters which reflect the effects of spin-orbit
interaction in our bi- and tri-layer MoS$_2$. Intriguingly, we estimate/speculate that in these multilayers the spin-orbit
splitting in the conduction band is quite large ($\sim$~50 meV) whereas the spin-orbit coupling parameter for the valence
band ($\sim$ 120...140meV) is somewhat small, as compared to the expected values for the MoS$_2$ monolayer.

The paper is organized as follows. The section~\ref{Theoretical_description} introduces the theoretical description of the
interlayer coupling and optically active transitions in 2H stacked multilayers. In the section~\ref{Experiment} we present
the experimental data for excitonic resonances in bi- and trilayer of MoS$_2$. In the section~\ref{Conclusion} we outline
the properties of multilayers and the possible applications of such materials. In the Appendix~\ref{Sec:0} the samples,
instrumentation and experimental details are presented. The Appendices~\ref{Sec:A} and \ref{Sec:B} contain the
derivation and discussion of the exciton $g$-factors in bi- and trilayer of MoS$_2$ respectively.

\section{Theoretical description}
\label{Theoretical_description}
We consider the optically active transitions at the $\mathrm{K}^\pm$ points in
$2$H-stacked multilayer S-TMD crystals encapsulated in hBN.
Our investigation is based on $\mathbf{k\cdot p}$ approximation. We focus on the optical properties of such
crystals at $\mathrm{K}^+$ point for brevity~\cite{molasNanoscale}. The results for $\mathrm{K}^-$ point can be
obtained by analogy.

\begin{figure}[t!]
		\centering
		\includegraphics[width=0.8\linewidth]{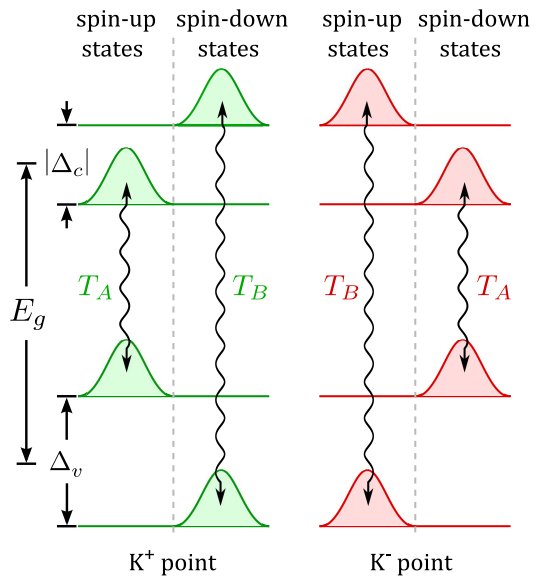}
		\caption{Sketch of band positions and optical transitions at the K$^+$ and K$^-$ points of the Brillouin zone
         in a monolayer of MoS$_2$. Green and red bump structures represent conduction and valence band states associated with
         transitions active correspondingly in the $\sigma^+$ and $\sigma^-$ polarizations. Solid wavy arrows denote all possible
         ($T_A$ and $T_B$) optical transitions. $|\Delta_c|$ and $\Delta_v$ define the absolute values of spin-orbit splitting
         in the CB and VB, respectively. $E_g$ is the single-particle band gap.}
		\label{fig:fig_0}
	\end{figure}

We briefly remind the features of S-TMD {\it monolayer} which will be used in the subsequent description of the
multilayer structures. Namely, a single layer crystal is a direct band gap semiconductor.
The maximum of valence (VB) and minimum of conduction (CB) bands are located at the $\mathrm{K}^\pm$ points of BZ.
Due to strong spin-orbit interaction both bands are spin-split (from hundreds
of $\mathrm{meV}$ in VB, up to tens $\mathrm{meV}$ in CB).
Hence, Bloch states in the corresponding points can be presented as a tensor product of spin and spinless states.
The spinless valence $|\Psi_v\rangle$ and conduction $|\Psi_c\rangle$ band states at the $\mathrm{K}^+$ point are made
predominantly from $d_{x^2-y^2}+id_{xy}$ and $d_{z^2}$ orbitals of transition metal atoms respectively
\cite{kormanyos,LiuElec2015}. Due to time-reversal symmetry (TRS) the analogous states at the $\mathrm{K}^-$ point
are complex conjugated to the previous ones, {\it i.e.} they are made from $d_{x^2-y^2}-id_{xy}$ and $d_{z^2}$ orbitals
accordingly. The TRS also dictates that Bloch states with the same band index ($c$ or $v$) but with opposite spins
in different valleys have equal energies. The crystal spatial symmetry together with TRS define the optical properties
of monolayers --- only $\sigma^\pm$ polarized light can be absorbed or emitted at the $\mathrm{K}^\pm$ points respectively.
Since the VB and CB are split, there are two possible optical transitions (which conserve spin) at the $\mathrm{K}^\pm$
points --- the lower-energy $T_A$ and the higher-energy $T_B$ ones. All the described features are depicted on
Fig.~\ref{fig:fig_0} for the case of MoS$_2$.

\begin{figure}
	\center{\includegraphics[scale=0.5]{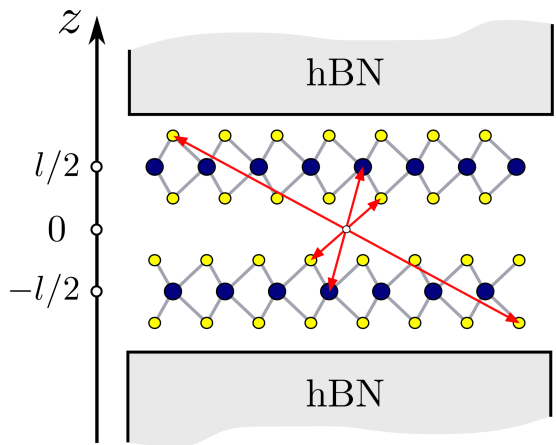}}
	\caption{
		The bilayer crystal embedded in between two thick hBN flakes, side view. The origin of all red
        arrows represents the inversion center of the crystal. The arrows, which lay on the same line, depict the inversion
        symmetry of the crystal. Namely, the tips of the corresponding arrows indicate the pair of atoms, which positions
        transform into each other after inversion procedure.}
        \label{fig:fig_bilayer_crystal}
\end{figure}

A {\it bilayer} crystal can be presented as two monolayers separated by a distance $l\sim 6$ {\AA}~\cite{Terrones2013},
and in which the top layer is rotated with respect to the bottom one by $180^\circ$. We arrange them in $z=l/2$ and
$z=-l/2$ planes, respectively (see Fig.~\ref{fig:fig_bilayer_crystal}). In this presentation, the crystal possesses
the  centrosymmetry $I$ with an inversion center in the $z=0$ plane. The bilayer embedded in between two flakes of the
same material (in our case, hBN), thickness and size conserves this symmetry. Therefore, one can extend the
$\mathbf{k\cdot p}$ model proposed in~\cite{gong_kp} also for this case.

Let us examine the Bloch states of the valence and conduction bands at the $\mathrm{K}^+$ point
of bilayer. We construct them from the Bloch states of top and bottom layers considering them separately.
Namely, we introduce the states $|\Psi^{(m)}_n\rangle\otimes|s\rangle$, where $m=1,2$ is a layer index
(for bottom and top layers, respectively), $n=v,c$ is a band index (for VB and CB) and $s=\uparrow,\downarrow$
specifies spin degree of freedom. The bottom (first) layer states $|\Psi^{(1)}_v\rangle$ and $|\Psi^{(1)}_c\rangle$
are made predominantly from $d_{x^2-y^2}+id_{xy}$ and $d_{z^2}$ orbitals of transition metal atoms
respectively~\cite{kormanyos,LiuElec2015}. They coincide with monolayer spinless states $|\Psi_v\rangle$ and $|\Psi_c\rangle$
mentioned before. The top (second) layer states $|\Psi^{(2)}_v\rangle$ and $|\Psi^{(2)}_c\rangle$ are made from
$d_{x^2-y^2}- id_{xy}$ and $d_{z^2}$ orbitals and coincide with spinless states at the
$\mathrm{K}^-$ point of monolayer. The top and bottom states are connected by the relation $|\Psi^{(2)}_n\rangle=K_0I|\Psi^{(1)}_n\rangle$,
where $K_0$  and $I$ are the complex conjugation and central inversion operators respectively. Finally, we suppose
the orthogonality of the basis states from different layers and bands $\langle\Psi_n^{(m)}|\Psi_{n'}^{(m')}\rangle=\delta_{nn'}\delta_{mm'}$.

The initial basis states in a given layer are affected by crystal fields of another layer and surrounding hBN medium.
Such fields being considered as a perturbation in $\mathbf{k\cdot p}$ model produce intra- and interlayer corrections
to the bilayer Hamiltonian. Namely, the intralayer ones renormalize the band gap $E_g$ and spin-splittings $\Delta_c, \Delta_v$
for the electron excitations in the considered layer. Due to the symmetry of the system, the parameters of the  other
layer get the same modifications. The interlayer corrections link the states from different layers. The symmetry
analysis of such terms demonstrates {\it i)} the strong coupling between VB basis states with the same spins;
{\it ii)} the quasi-momentum dependent coupling between CB basis states with the same spins; {\it iii)} the coupling
between VB and CB basis states of the opposite spins, which appears due to spin-orbit interaction. The latter term is
supposed to be small and is omitted from our study. In this case, the spin-up and spin-down states of bilayer are
decoupled and can be considered separately.

As a result, the VB Hamiltonian
written in the basis $\{|\Psi^{(1)}_v\rangle\otimes|s\rangle,|\Psi^{(2)}_v\rangle\otimes|s\rangle\}$ takes the form
\begin{equation}
H^{(2)}_{vs}=\left[
\begin{array}{cc}
\sigma_s\frac{\Delta_v}{2} & t  \\
t & -\sigma_s\frac{\Delta_v}{2}  \\
\end{array}
\right],
\end{equation}
where $\sigma_s=+1(-1)$ for $s=\uparrow(\downarrow)$.
The parameter $t\sim 40-70$ \text{meV}~\cite{gong_kp} defines the coupling between valence bands
from different layers. The CB Hamiltonian, written in the basis $\{|\Psi^{(1)}_c\rangle\otimes|s\rangle,|\Psi^{(2)}_c\rangle\otimes|s\rangle\}$
is
\begin{equation}
H^{(2)}_{cs}=\left[
\begin{array}{cc}
E_g+\sigma_s\frac{\Delta_c}{2} & uk_+ \\
uk_- & E_g-\sigma_s\frac{\Delta_c}{2} \\
\end{array}
\right],
\end{equation}
where $E_g$ is the band gap of bilayer and $k_\pm=k_x\pm ik_y$. Both Hamiltonians are written up to $O(k^2)$.

The bilayer VB have the energies $E^\pm_v=\pm\sqrt{\Delta_v^2/4+t^2}$. The corresponding
upper-energy eigenstates are
\begin{align}
&|\Phi^+_{v\uparrow}\rangle=\Big[\cos\theta|\Psi^{(1)}_v\rangle + \sin\theta|\Psi^{(2)}_v\rangle\Big]\otimes|\uparrow\rangle,
\nonumber \\
&|\Phi^+_{v\downarrow}\rangle=\Big[\sin\theta|\Psi^{(1)}_v\rangle+
\cos\theta|\Psi^{(2)}_v\rangle\Big]\otimes|\downarrow\rangle,
\end{align}
where we introduced $\cos(2\theta)=\Delta_v/\sqrt{\Delta_v^2+4t^2}$.
The low-energy eigenstates $|\Phi^-_{v\uparrow}\rangle$ and $|\Phi^-_{v\downarrow}\rangle$
can be derived from the first ones by replacing $\cos\theta\rightarrow -\sin\theta, \sin\theta\rightarrow \cos\theta$.
The new VB are doubly degenerated by spin.

The CB states do not interact with each other in $\mathrm{K}^+$ valley ({\it i.e.} in $k_x,k_y\rightarrow 0$ limit).
Hence, in leading order they are not mixed and form doubly degenerated bands with energies $E_g\pm\Delta_c/2$.
In further we focus on MoS$_2$ bilayer with $\Delta_c<0$. For this case, the upper and lower energy CB states
are $\{|\Psi^{(1)}_c\rangle\otimes|\downarrow\rangle,
|\Psi^{(2)}_c\rangle\otimes|\uparrow\rangle\}$ and
$\{|\Psi^{(1)}_c\rangle\otimes|\uparrow\rangle,
|\Psi^{(2)}_c\rangle\otimes|\downarrow\rangle\}$, respectively.

\begin{figure}
	\center{\includegraphics[scale=0.45]{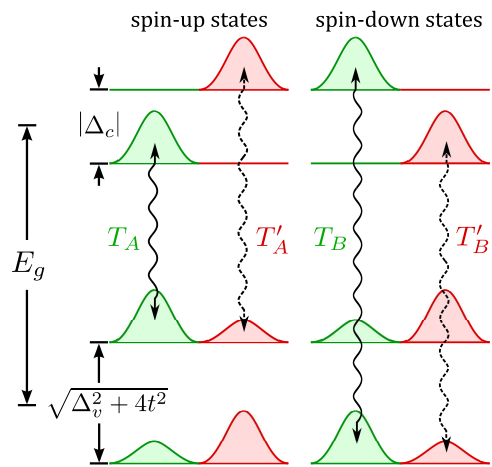}}
	\caption{\label{fig:fig_3}
		Sketch of the bands positions and optical transitions at the K$^+$ of the BZ in a bilayer of MoS$_2$.
Green and red bump structures represent conduction and valence band states associated with transitions active
in the $\sigma^+$ and $\sigma^-$ polarizations respectively. Solid (dashed) wavy arrows denote optical
transitions due to the intralayer (interlayer) A and B excitons. $|\Delta_c|$ and $\sqrt{\Delta^2_v+4t^2}$
denote the splitting in the CB and VB, respectively. $E_g$ is the single particle band gap.}
\end{figure}

All new energy states of bilayer system are depicted graphically in Fig.~\ref{fig:fig_3}. We
divide them on spin-up (left) and spin-down (right) subsets for clarity. The single and
doubled bumps in the figure represent the new CB and VB states. The size of the bump describes
the probability to observe the new composite quasiparticle in the bottom (green part)
or the top (red part) layers. Note that the valence band states can be found in both layers.
That makes these states optically active in both polarizations. Namely, the transitions from
any valence band state into $|\Psi^{(1)}_c\rangle$ and $|\Psi^{(2)}_c\rangle$ states are possible
in $\sigma^+$ and $\sigma^-$ circularly polarised light, respectively.
In consequence, the system demonstrates four types of optically active exciton resonances --- two intralayer
and two interlayer ones. The half of all transitions --- $T_{A}(T_B)$ between spin-up(spin-down) bands are
presented in Fig.~\ref{fig:fig_3}.

There are two intralayer exciton transitions in bilayer
\begin{align}
T_A:\left\{\begin{array}{cc}
|\Phi^+_{v\uparrow}\rangle\rightarrow|\Psi^{(1)}_c\rangle\otimes|\uparrow\rangle, \\
|\Phi^+_{v\downarrow}\rangle\rightarrow|\Psi^{(2)}_c\rangle\otimes|\downarrow\rangle;
\end{array}\right.
\end{align}
\begin{align}
T_B:\left\{\begin{array}{cc} |\Phi^-_{v\uparrow}\rangle\rightarrow|\Psi^{(2)}_c\rangle\otimes|\uparrow\rangle, \\
|\Phi^-_{v\downarrow}\rangle\rightarrow|\Psi^{(1)}_c\rangle\otimes|\downarrow\rangle.
\end{array}\right.
\end{align}
All of them have the same intensity $I=I_0\cos^2\theta$, where $I_0$ is the intensity of exciton
line in monolayer.

There are two interlayer exciton transitions in bilayer
\begin{align}
T_A':\left\{\begin{array}{cc}|\Phi^+_{v\uparrow}\rangle\rightarrow|\Psi^{(2)}_c\rangle\otimes|\uparrow\rangle, \\
|\Phi^+_{v\downarrow}\rangle\rightarrow|\Psi^{(1)}_c\rangle\otimes|\downarrow\rangle;
\end{array}\right.
\end{align}
\begin{align}
T_B':\left\{\begin{array}{cc}|\Phi^-_{v\uparrow}\rangle\rightarrow|\Psi^{(1)}_c\rangle\otimes|\uparrow\rangle, \\
|\Phi^-_{v\downarrow}\rangle\rightarrow|\Psi^{(2)}_c\rangle\otimes|\downarrow\rangle.
\end{array}\right.
\end{align}
They have the intensities  $I'=I_0\sin^2\theta$.
Finally, since the intra- and inter-layer exciton transitions are active in opposite circular polarization
at a given $\mathrm{K}$ point, they should have the opposite signs of $g$-factors (see Appendix~\ref{Sec:A}).

\begin{figure}
	\center{\includegraphics[scale=0.5]{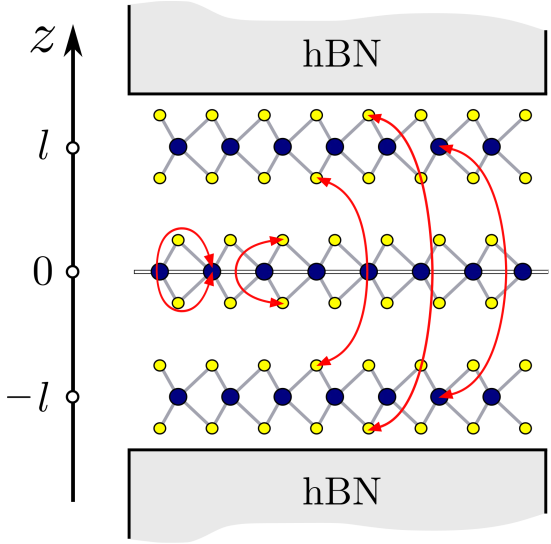}}
	\caption{\label{fig:fig_trilayer_crystal}
		The trilayer crystal embedded in between two thick hBN flakes, side view. The mirror symmetry plane $z=0$ is
        represented by thin rectangular. The ends of red curvy arrows couple the atoms, which coordinates transform to
        each other after mirror symmetry transformation operation.}
\end{figure}

For a {\it trilayer} case, a crystal can be presented as three monolayers each separated
by distance $l$, with the middle layer is $180^\circ$ rotated around the external ones (see Fig.~\ref{fig:fig_trilayer_crystal}).
Similarly to the bilayer case, we focus on properties of quasiparticles
at the $\mathrm{K}^+$ point. The basis states of a trilayer are
$\{|\Psi^{(1)}_n\rangle\otimes|s\rangle, |\Psi^{(2)}_n\rangle\otimes|s\rangle, |\Psi^{(3)}_n\rangle\otimes|s\rangle\}$.
They belong to $z=-l$, $z=0$ and $z=l$ layers, respectively. In this case, the crystal
has mirror symmetry $\sigma_h$, with the mirror plane $z=0$. It determines the symmetry
relations $|\Psi^{(3)}_n\rangle=\sigma_h|\Psi^{(1)}_n\rangle, |\Psi^{(2)}_n\rangle=\sigma_h|\Psi^{(2)}_n\rangle$.
The properties of $|\Psi^{(1)}_n\rangle$  and $|\Psi^{(2)}_n\rangle$ states are the
same as in bilayer discussed in the main text. The properties of $|\Psi^{(3)}_n\rangle$ states
coincide with $|\Psi^{(1)}_n\rangle$ according to the crystal symmetry.
Hence, in $\mathrm{K}^+$ point, the $1$-st and $3$-d layers absorb only $\sigma^+$ polarised light,
while the $2$-nd one is active in $\sigma_-$ polarization. Note that the trilayer
encapsulated in between to equal hBN flakes conserves the mirror symmetry.

\begin{figure*}[t]
		\includegraphics[width=0.8\linewidth]{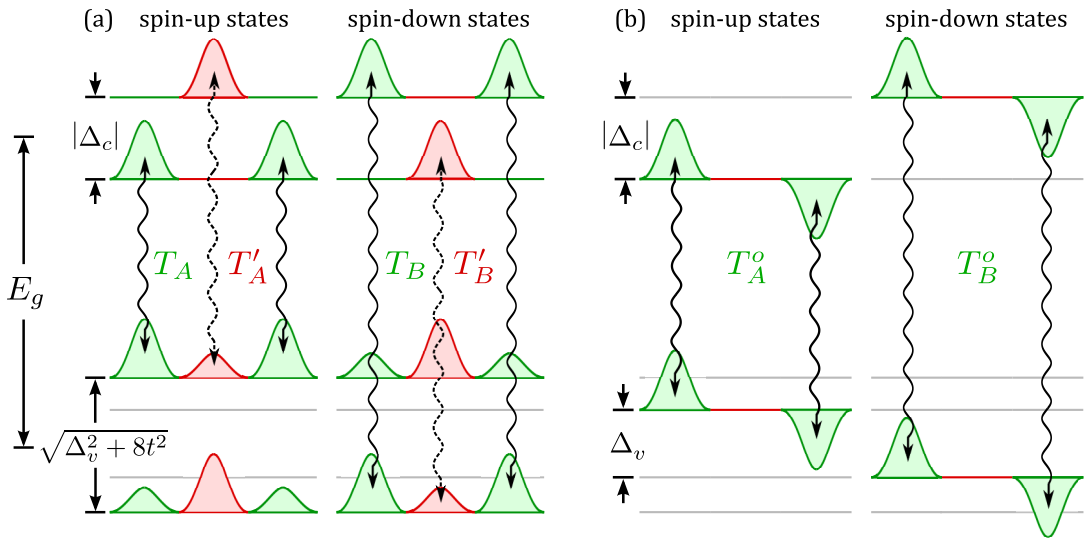}%
		\caption{\label{fig:fig_1A}
		Sketch of bands positions and optical transitions between (a) even and (b) odd states at the $\mathrm{K}^+$
        of the BZ in a bilayer of MoS$_2$. Green and red bump structures represent conduction and valence band states
        associated with transitions active correspondingly in the $\sigma^+$ and $\sigma^-$ polarizations. Solid (dashed)
        wavy arrows denote optical transitions due to the intralayer (interlayer) A and B excitons. $|\Delta_c|$ and
        $\sqrt{\Delta^2_v+8t^2}$ ($\Delta_v$) denote the splitting in the CB and even (odd) states of VB, respectively.
        $E_g$ is the single particle band gap. }
	\end{figure*}

We derive the effective $\mathbf{k\cdot p}$ Hamiltonian for
considering system within several approximations: {\it i)}
the interlayer spin-orbit coupling between CB and VB states with opposite spins from different
layers is neglected, like in bilayer; {\it ii)}
the intralayer crystal field corrections are equal for each layer,
{\it i.e.} the bands of each layer are characterised by the same band-gap $E_g$ and
spin-splitting $\Delta_v,\Delta_c$ parameters, the magnitudes of which, however,
can deviate from their bilayer analogs; {\it iii)} we neglect the coupling
between the states of the $1$-st and $3$-d layers. In this approach, like in bilayer case,
the spin-up and spin-down states of trilayer are decoupled.
The Hamiltonian for VB states, written in the basis
$\{|\Psi^{(1)}_v\rangle\otimes|s\rangle,|\Psi^{(2)}_v\rangle\otimes|s\rangle,
|\Psi^{(3)}_v\rangle\otimes|s\rangle\}$, is
\begin{equation}
H^{(3)}_{vs}=\left[
\begin{array}{ccc}
\sigma_s\frac{\Delta_v}{2} & t & 0  \\
t & -\sigma_s\frac{\Delta_v}{2} & t  \\
0 & t & \sigma_s\frac{\Delta_v}{2}  \\
\end{array}
\right].
\end{equation}
The Hamiltonian for conduction bands, written in the basis
$\{|\Psi^{(1)}_c\rangle\otimes|s\rangle,|\Psi^{(2)}_c\rangle\otimes|s\rangle,|\Psi^{(3)}_c\rangle\otimes|s\rangle\}$, reads
\begin{equation}
H^{(3)}_{cs}=\left[
\begin{array}{ccc}
E_g+\sigma_s\frac{\Delta_c}{2} & uk_+ & 0 \\
uk_- & E_g-\sigma_s\frac{\Delta_c}{2} & uk_- \\
0& uk_+&  E_g+\sigma_s\frac{\Delta_c}{2}\\
\end{array}
\right].
\end{equation}
The system possesses a mirror symmetry. Hence, it is convenient to introduce the new basis states
which are even
$|\Upsilon^{(1)}_n\rangle=\frac{1}{\sqrt{2}}\big[|\Psi^{(1)}_n\rangle + |\Psi^{(3)}_n\rangle\big]$,
$|\Upsilon^{(2)}_n\rangle=|\Psi^{(2)}_n\rangle$ and odd
$|\Upsilon^{(3)}_n\rangle=\frac{1}{\sqrt{2}}\big[|\Psi^{(1)}_n\rangle - |\Psi^{(3)}_n\rangle\big]$
under $\sigma_h$ transformation.
The trilayer Hamiltonians have a block-diagonal form in the basis
$\{|\Upsilon^{(1)}_n\rangle\otimes|s\rangle,
|\Upsilon^{(2)}_n\rangle\otimes|s\rangle, |\Upsilon^{(3)}_n\rangle\otimes|s\rangle\}$
\begin{equation}
\mathcal{H}^{(3)}_{vs}=\left[
\begin{array}{ccc}
\sigma_s\frac{\Delta_v}{2} & \sqrt{2}t & 0  \\
\sqrt{2}t & -\sigma_s\frac{\Delta_v}{2} & 0  \\
0 & 0 & \sigma_s\frac{\Delta_v}{2}  \\
\end{array}
\right],
\end{equation}
\begin{equation}
\mathcal{H}^{(3)}_{cs}=\left[
\begin{array}{ccc}
E_g+\sigma_s\frac{\Delta_c}{2} & \sqrt{2}uk_+ & 0 \\
\sqrt{2}uk_- & E_g-\sigma_s\frac{\Delta_c}{2} & 0 \\
0 & 0 &  E_g+\sigma_s\frac{\Delta_c}{2}\\
\end{array}
\right].
\end{equation}
The $2\times 2$ blocks have the structure of bilayer Hamiltonians. Therefore, all the properties
of the trilayer even states can be obtained from bilayer result by replacing $t\rightarrow \sqrt{2}t, u\rightarrow \sqrt{2}u$.
Consequently, there are doubly degenerate valence and conduction bands, which define the
intensive $T_A$ and $T_B$ and weak $T_A'$ and $T_B'$ groups excitonic transitions,
which are active in opposite polarizations of light (see Fig.~\ref{fig:fig_1A}(a)).
The $1\times 1$ block has the structure of monolayer Hamiltonian. Hence, there are only
$T^{o}_A$ and $T^{o}_B$ transitions between odd states of trilayer. They are active in the
same polarization as $T_A$ and $T_B$ transitions of monolayer (see Fig.~\ref{fig:fig_1A}(b)).
Summarising, the trilayer possesses two types of ''even'' intralayer excitons --- $T_A$ and $T_B$,
two types of ''odd''intralayer ones --- $T^{o}_A$ and $T^{o}_B$. All of them are active in $\sigma^+$
polarization at the K$^+$ point and have the $g$-factors of the same sign. The interlayer ''even''
excitons --- $T_A'$ and $T_B'$, conversely are active in $\sigma^-$ polarization at the K$^+$ point
and have an opposite sign to intralayer excitons $g$-factors. The detailed description of trilayer's $g$-factors
is presented in Appendix~\ref{Sec:B}.

Note that the current theoretical description predicts the number of exciton resonances and their polarization properties.
However, the relative position of $T_A$ and $T^o_A$ exciton remains an open question. Indeed, both lines are active in the
same polarization and have similar $g$-factors, that makes impossible to identify each of them from the experiment. The
analysis of the interference effects together with the consideration of the Coulomb interaction between new quasiparticles
answers this question, which is, however, beyond the scope of the current paper. In our case, we took the position of exciton
resonances following the position of the valence and conduction bands, derived in single-particle $\mathbf{k\cdot p}$
approximation, for definiteness.

\begin{figure}[t]
		\includegraphics[width=0.8\linewidth]{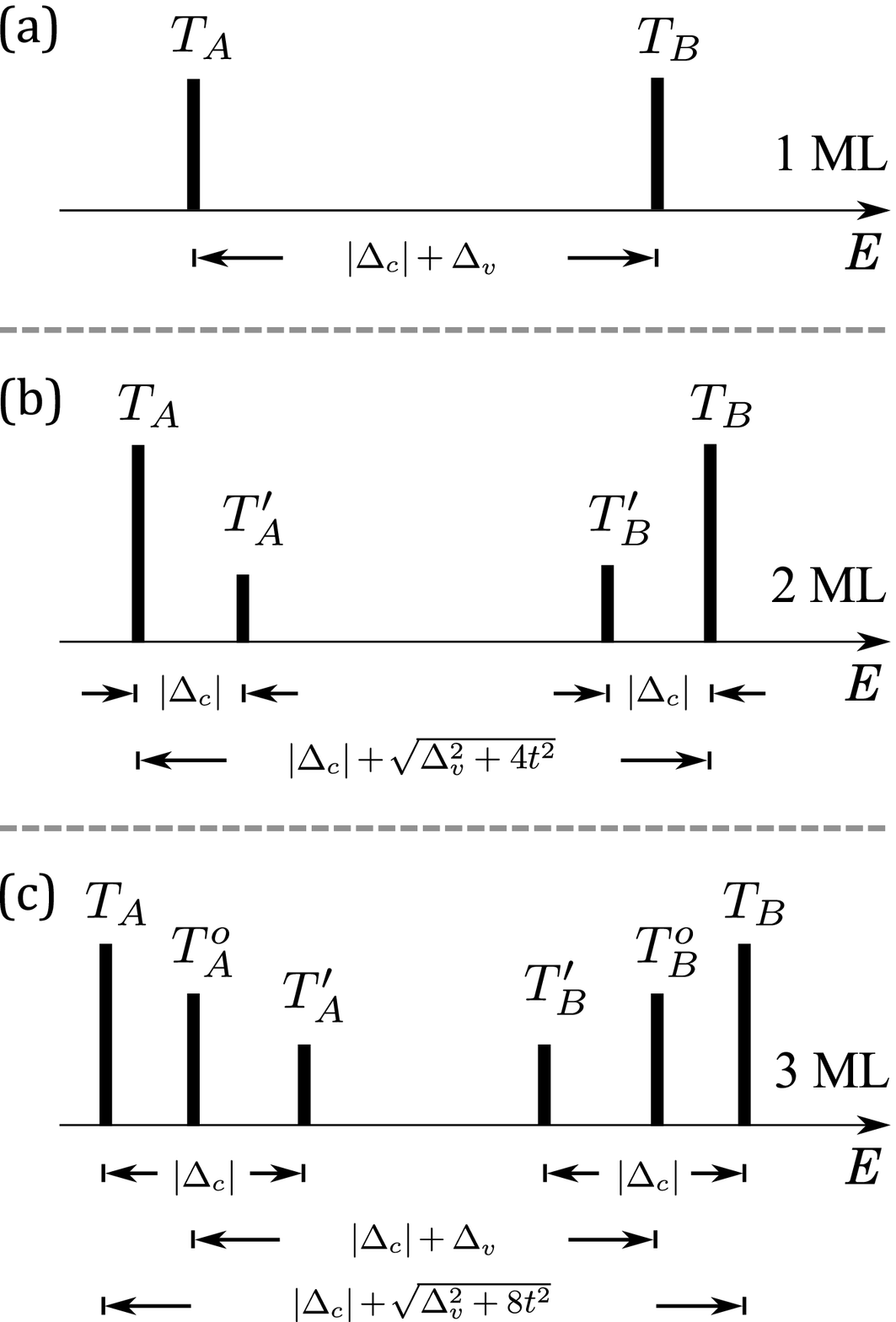}%
		\caption{\label{fig:fig_transitions}
		Sketch of optical transitions in (a) mono-, (b) bi- and (c) trilayer S-TMD crystals. The position and size of black
        thick segments represent the energies and intensities of the corresponding transitions. The values of parameters
        $\Delta_c$ and $\Delta_v$ for bilayer and trilayer can deviate from their monolayer analogs.}
	\end{figure}

Our single-particle considerations are summarized in Fig.~\ref{fig:fig_transitions}. This picture exhibits the position and
number of optically active transitions in mono-, bi- and trilayer S-TMD crystals.

\section{Absorption resonances in MoS$_{2}$ multilayers}
\label{Experiment}
\begin{figure}[t!]
		\centering
		\includegraphics[width=0.8\linewidth]{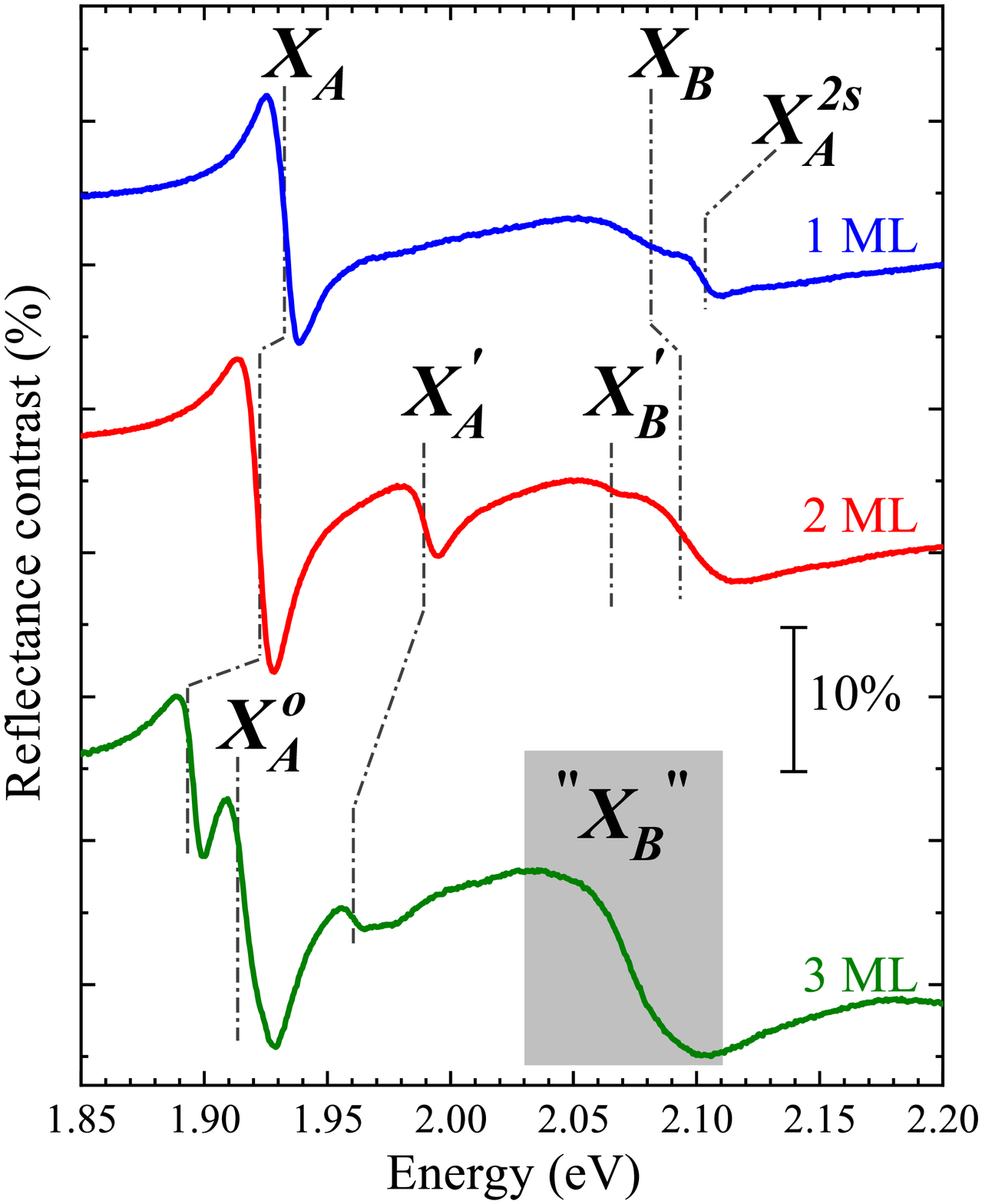}%
		\caption{Reflectance contrast spectra of MoS$_2$ layers encapsulated in hBN measured at $T$=5~K. The spectra are vertically
        shifted for clarity.}
		\label{fig:fig_1}
	\end{figure}
\begin{figure*}[t!]
		\centering
		\includegraphics[width=0.8\linewidth]{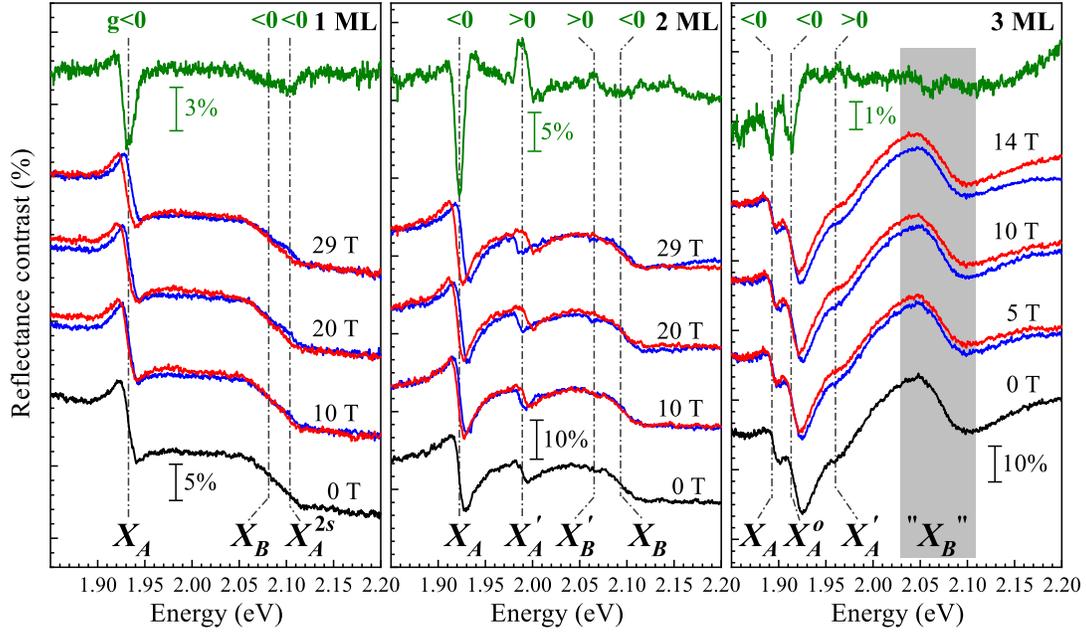}%
		\caption{Helicity-resolved reflectance contrast spectra of MoS$_2$ layers encapsulated in hBN for selected values of magnetic
        field measured at $T$=4.2~K. The red and blue curves correspond to $\sigma^+$ polarization of reflected light in $B$ and $-B$
        configurations of magnetic field (applied perpendicularly to the layers plane), respectively. The spectra are vertically
        shifted for clarity.}
		\label{fig:fig_2}
	\end{figure*}
\begin{figure*}[t!]
		\centering
		\includegraphics[width=0.8\linewidth]{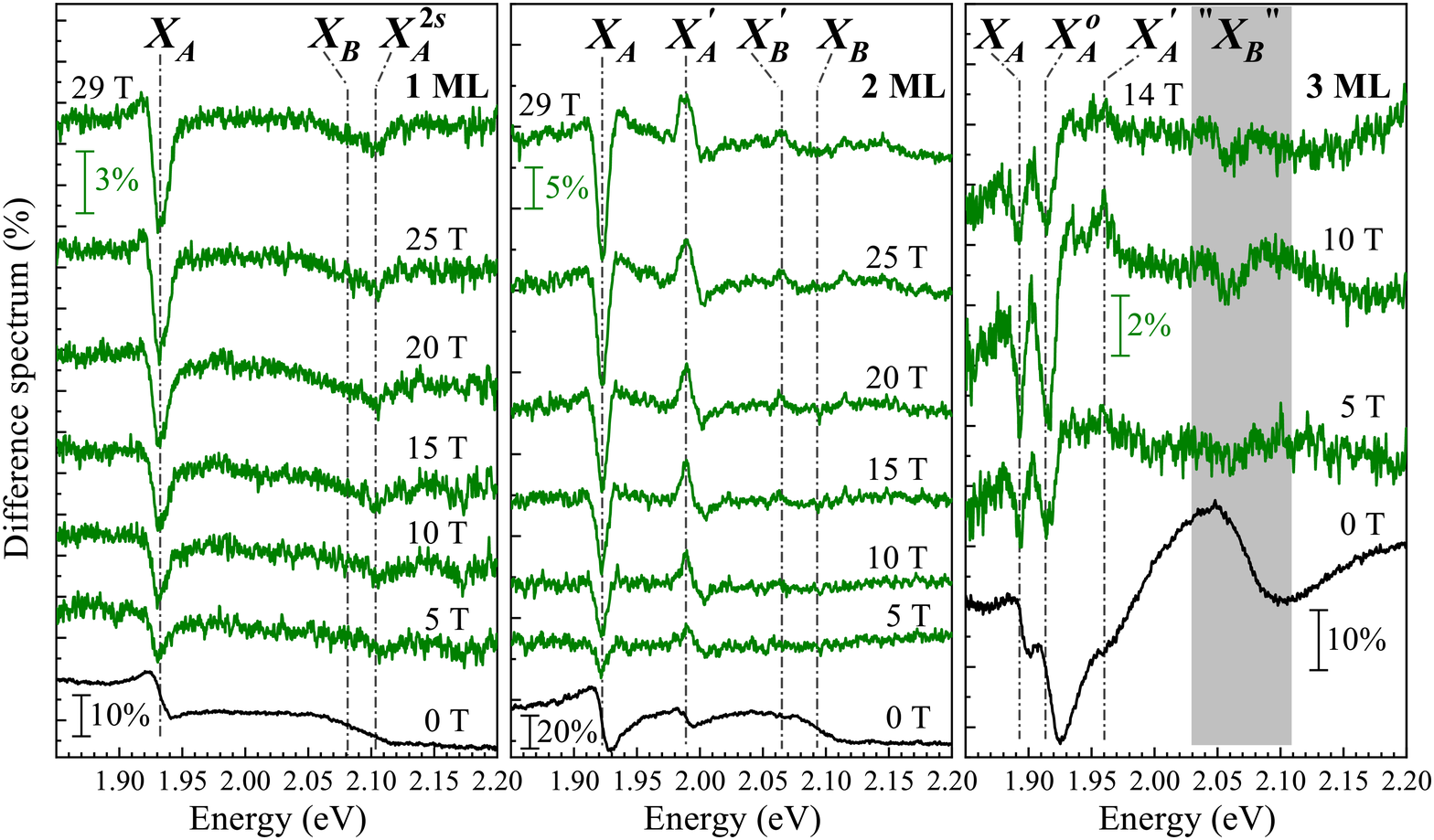}%
		\caption{Zero-field RC spectra (black) with difference spectra $[RC(\sigma^+,B)-RC(\sigma^+,-B)]$ at selected values of $B$
         field for 1ML (left panel), 2ML (middle panel) and 3ML (right panel) MoS$_2$ encapsulated in hBN. A deep in the difference
         spectra indicate a negative $g$-factor, while a peak indicates a positive $g$-factor.}
		\label{fig:fig_exp_3}
	\end{figure*}
In the experiments, we have investigated a set of van der Waals heterostructures with optically active parts consisting of MoS$_2$ mono-
(1ML), bi- (2ML) and trilayer (3ML). The structures were prepared on ultraflat unoxidized silicon substrates by combination of
exfoliation and dry transfer techniques \cite{gomez}. In order to achieve the highest possible quality of our samples, as well as to
preserve the characteristic inverse and mirror symmetries of, respectively, 2H-stacked bi- and trilayer, the microcleaved MoS$_2$ flakes
were sandwiched between two hBN flakes exfoliated from high-quality bulk crystals grown under high-pressure conditions \cite{taniguchi2007}.
Further information on the preparation of samples and their characterization is provided in
Appendix~\ref{Sec:0}.

Experiments consisted of measurements of the reflectance contrast spectra, which are defined as
\begin{equation}
RC(E)=\frac{R(E)-R_0(E)}{R(E)+R_0(E)}\times100\%,
\end{equation}
where $R(E)$ is the reflectance spectrum when the light is focused on the MoS$_2$ flake and $R_0(E)$
is the reflectance from the region outside the flake. Two different micro-optical setups have been used
for measurements: one setup for measurements in the absence of magnetic fields and the second one for
 measurements in magnetic fields supported by either a 14\,T superconducting magnet or a 29\,T resistive
 magnet. A spatial resolution of about $1\,\mu m$ (light spot diameter) is characteristic of our
 free-beam-optics setup used for experiments at zero magnetic field. Instead, the fiber-optics-based
 arrangement applied for experiments in magnetic fields provides a poorer spatial resolution, of about
 $10\,\mu m$. A spectral resolution of 0.32\,nm has been assured for both setups. A combination of a
 quarter wave plate and a polarizer are used to analyse the circular polarization of signals. The
 measurements are performed with a fixed circular polarization, whereas reversing the direction of
 magnetic field yields the information corresponding to the other polarization component due to
 time-reversal symmetry. More on experimental details can be found in Appendix~\ref{Sec:0}.

The collection of low temperature $(\sim 5\mathrm{K})$ RC spectra using our $B=0$ setup, measured in the
spectral range of the optical gap (onset of strong absorption) of the MoS$_2$ layers is presented in
Fig.~\ref{fig:fig_1}. The observed, more or less pronounced resonances, show the dispersive spectral shape,
correspond to typically, strongly bound direct bandgap excitons in S-TMD layers. Although our theoretical
considerations neglect the effects of Coulomb binding and account only for interband transitions, a distinct
resemblance between the measured spectral evolution (Fig.~\ref{fig:fig_1}) and that theoretically concluded
in Fig.~\ref{fig:fig_transitions} can be recognized.

In the following we assume that each interband transition $T_X$ is associated with the corresponding excitonic
resonance $X$ and we proceed with the assignment of the resonances observed in the experiment. We suppose that the
resonating excitons are shifted in energy, by their binding energies, with respect to the associated interband
transitions: $E(X)=E(T_X)-E_b(X)$. Binding energies might be, however, different for different excitons and this
fact must be taken into account when comparing the energy position of $X$-resonances and $T_X$-transitions.
Important for the assignment of the observed resonances are the results of the polarization resolved measurements
carried out in magnetic fields (see Figs.~\ref{fig:fig_2} and \ref{fig:fig_exp_3}). As discussed above, we expect
that our multilayers host two different types of intra- and interlayer excitons, each of them being distinguishable
by their different polarization properties ($g$-factors of opposite sign and various magnitudes). At this point we must
admit that all subtle spectral features which are well visible in the RC spectra measured at $B=0$ with high spatial
resolution are somewhat less pronounced in the magneto-optical measurements which imply worse spatial resolution.
Certain degree of sample inhomogeneity is an obvious cause of this drawback. In consequence, in the case of weak
and/or broad resonances the information about their $g$-factors is not easily extractable from the raw magneto-RC data.
This information becomes more apparent if we inspect the RC-polarization spectra. These spectra have been constructed
as a difference between the RC-spectra measured at the same strength but for two opposite direction of the magnetic
field (that mimic, due to the time reversal symmetry, the spectra corresponding to $\sigma^+$ and $\sigma^-$ circular
polarization of the reflected light~\cite{arora2017,koperskiMagneto}). As can be deduced from the data shown in
Figs.~\ref{fig:fig_2} and \ref{fig:fig_exp_3}, the apparent dips in our RC polarization spectra correspond to
excitonic resonances with negative $g$-factors whereas the characteristic upswings mark the resonances with
positive $g$-factors.

Very first classification of the observed resonance takes into account a relatively large spin-orbit splitting in the
MoS$_2$ valence band, rising two groups of excitons associated with well-separated upper ($A$-excitons) and lower
($B$-excitons) valence band subbands. The 1ML spectrum shown in Fig.~\ref{fig:fig_1}, resembles that previously reported
for a high quality 1ML MoS$_2$ encapsulated in hBN~\cite{cadizMoS2,robert2018}. It depicts three resonances: one well-separated
resonance due $A$-exciton (ground state), $X_A$, and two other superimposed resonances, due to $B$-exciton ($X_B$) and
the first excited (2s) state of $A$-exciton ($X_A^{2s}$). $X_A$ and $X_B$ resonances are associated with $T_A$ and $T_B$
transitions sketched in Fig.~\ref{fig:fig_0}; the excited excitonic states are beyond the frame of our single particle
theoretical approach. As expected, the $X_A$, $X_B$ and $X_A^{2s}$ excitonic resonances display similar polarization
properties in the magneto-optical experiments (see Fig.~\ref{fig:fig_2}). We estimate $g_{X_A} \approx -4$ in agreement
with previous reports~\cite{mitioglu2016,stierNatCom,wu2018}; the accurate estimation of the amplitudes of $g_{X_B}$
and $g_{X_A^{2s}}$ is more cumbersome though both these values are also negative.

The RC spectrum of the 2ML (Fig.~\ref{fig:fig_1}) shows 4 resonances~\cite{horng2018} which, in reference to our
theoretical expectations (see Figs.~\ref{fig:fig_3} and \ref{fig:fig_transitions}), are assigned to the pair of
intra- $(X_A)$ and interlayer $(X_A')$ $A$-excitons, and to the analogous pair of intra- $(X_B)$ and interlayer
$(X_B')$ $B$-excitons. We use the signs and magnitude of the exciton $g$-factors as fingerprints the intra- and interlayer
nature of the $X_A$- and $X_A'$-excitons respectively~\cite{arora2017}. Namely, we conclude that $g_{X_A}\approx-4$,
while $g_{X_A'}\approx8$. Again, the exact values for the $g$-factor amplitudes of $B$-excitons are hard to be
precisely estimated. Nonetheless it is rather clear that $g_{X_B}$ is negative whereas $g_{X_B'}$ is positive.
Whereas our theoretical model meets the experimental data at the qualitative level, the energy ladder of the observed
resonances requests further comments. Characteristic for the theoretical modelling is the fact that the energy distance,
$\Delta(T_A',T_A)$, between the $T_A'$ and $T_A$ transitions is the same as the energy separation, $\Delta(T_B,T_B')$
between the $T_B$ and $T_B'$ transitions, both differences being determined by the amplitude $\Delta_c$ of the spin
orbit splitting in the conduction band of the 2ML: $\Delta(T_A',T_A)=\Delta(T_B,T_B')=|\Delta_c|$. This property is
not seen in the experiment: we estimate that $\Delta(X_A',X_A)\simeq70$\,meV, whereas $\Delta(X_B,X_B')\simeq30$\,meV.
The inconsistency between theory and experiment is likely due to different binding energies of excitons associated with
different interband transitions: $E(X)=E(T_X)-E_b(X)$. In the first approximation we assume that binding energies of our
inter- and intralayer excitons are indeed different but that the excitons within each pair of indirect and direct
resonances are the same: $E_b(intra)=E_b(X_A)=E_b(X_B)\neq E_b(X_A')=E_b(X_B')=E_b(inter)$. Then, comparing the
theoretical prediction with the experimental data one obtains that
$|\Delta_c|=\Delta(T_A',T_A)=\Delta(T_B,T_B')=[\Delta(X_A',X_A)+\Delta(X_B,X_B')]/2\simeq50$\,meV and that
$E_b(intra)=E_b(inter)+20$\,meV. Larger binding energies of intralayer excitons with respect to those of
interlayer excitons are logically expected since Coulomb attraction should be indeed stronger/weaker when the electron
and hole charges are localized in the same or in the neighboring layers, as for the case of intra- or interlayer excitons,
respectively. At first sight, the estimated amplitude of the spin orbit splitting in the conduction band of the MoS$_2$
bilayer appears to be surprisingly large. In the case of 1L MoS$_2$, the commonly accepted results of the DFT calculations
predict $\Delta_c$ in the range of few meV, though recent experimental works point out towards much higher values of about
15\,meV \cite{Pisoni2018}. According to our theoretical approach, the spin orbit interaction is sensitive to interlayer
coupling and it is also affected by the interaction with the surrounding material (hBN). Thus the amplitudes of the
$\Delta_c$ as well as $\Delta_v$ parameters are expected to evolve with the numbers of the stacked MoS$_2$ layers.
Keeping our assumption about equal binding energies for $X_A$ and $X_B$ excitons we note that the expected energy
separation between these excitonic resonance is given by $|\Delta_c|+\sqrt{\Delta_v^2+4t^2}$ (see Figs.~\ref{fig:fig_3}
and \ref{fig:fig_transitions}), to be compared to the value of 170\,meV estimated from the experiment. With the theoretically
estimated value of $t\simeq40$\,meV \cite{gong_kp} and the derived above $|\Delta_c|\simeq50$\,meV we find $\Delta_v\simeq90$\,meV.
Visibly, however, the ``effective'' spin orbit splitting in the valence band of the MoS$_2$ bilayer is larger, as given
by $\sqrt{\Delta_v^2+4t^2}\simeq120$\,meV (see Fig.~\ref{fig:fig_3}).

Focusing now on the trilayer spectra (see bottom panel of Fig.~\ref{fig:fig_1}) we observe a characteristic triplet of
$A$-exciton resonances, which is in accordance with our theoretical expectations (see Fig.~\ref{fig:fig_transitions}c).
As expected, two strong $X_A$ and $X_A^o$ resonances are characterised by negative $g$-factors whereas the $g$-factor of
the interlayer $X_A'$ exciton is pretty much positive. The theoretically anticipated triplet structure of the
$''X''_B$-exciton is not resolved in the experiment but likely hidden within the observed broad spectrum of the $B$-exciton.
Even with unresolved fine structure of the $B$-exciton for our trilayer MoS$_2$, a rough estimation of $\Delta_c$ and $\Delta_v$
parameters can be done. Expecting that the energy separation between $X_A'$ and $X_A$ is $\Delta(X_A',X_A)=|\Delta_c| +E_b(X_A)-E_b(X_A')$
and reading $\Delta(X_A',X_A)\simeq70$\,meV from the experiment we conclude that $|\Delta_c|\simeq50$\,meV if the difference
between binding energies of the intralayer $X_A$ and interlayer $X_A'$ excitons is  $E_b(X_A)-E_b(X_A')\simeq20$\,meV, {\it i.e.}
the same as we found for intra- and interlayer excitons in 2ML MoS$_2$. On the other hand, supposing that binding energies of
two $X_A$ and $X_A^o$ intralayer excitons are the same we expect that they are separated in energy by
$\Delta(X_A^o,X_A)=(\sqrt{\Delta_v^2+8t^2}-\Delta_v)/2$ and extract $\Delta_v\simeq140$\,meV when reading
$\Delta(X_A^o,X_A)\simeq20$\,meV from the experiments and assuming again that $t\simeq40$\,meV. With the above estimation,
the band-edge structure of the 3L MoS$_2$ at the $\mathrm{K}^\pm$ points of the Brillouin zone is concluded to consist of two
conduction band subbands split by 50\,meV and 4 valence band subbands with outer two subbands split by 180\,meV and the inner
two subbands split by 140\,meV.

\section{Conclusion}
\label{Conclusion}
We have performed magnetooptical $\mu$-reflectance measurements along with $\mathbf{k\cdot p}$ theory based modelling on few-layer
MoS$_{2}$ encapsulated in hBN structures, revealing the intralayer and interlayer nature of the newly discovered transitions. Such
resonances form due to hybridisation of valence and conduction bands states when one adds new layers to the system.

The experiment {\it i)} manifests the new exciton resonances and their number in bi- and trilayers; {\it ii)} demonstrates that
$g$-factors of intralayer and interlayer transitions have opposite signs; {\it iii)} the $g$-factor values of the latter ones are
much larger (by magnitude) than of the former ones.

The symmetry based $\mathbf{k\cdot p}$ description of the bi- and trilayers gives the natural explanation of the aforementioned
experimental observations. Moreover, from the general symmetry point of view the theoretical model also predicts {\it i)} the
renormalization of the band gap $E_g$ and spin splittings $\Delta_c, \Delta_v$; {\it ii)} the deviation of the bi- and trilayer
exciton $g$-factors from their monolayer analogs. These two phenomena appear as a result of the coupling between different layer
of the system and effects of dielectric screening induced by hBN.

Finally, we point out some unique properties of the exciton states of the trilayer.
First, we mention the existence of two groups of exciton resonances --- ``even'' and ``odd'', which do not interact with each other
because they belong to different irreps of the in-plane mirror symmetry of the system. However, electric field, applied
perpendicularly to the crystal's plane, violates this symmetry and causes the controllable coupling of these states. Such feature can be
used in future exciton based applications. We also note that the electrical positive and negative charges of interlayer excitons are
separated between different layers. Such a property can be interesting in photovoltaic applications of S-TMD systems.

\section*{Acknowledgements}

The work has been supported by
the ATOMOPTO project (TEAM programme of the Foundation for Poli
sh Science co-financed by the EU within the ERDFund),
the EC Graphene Flagship project (no. 604391),
the National Science Centre (grants no. DEC-2013/10/M/ST3/00791, UMO-2017/24/C/ST3/00119), the Nanofab facility of the Institut N\'eel,
CNRS UGA and LNCMI-CNRS, a member of the European Magnetic Field Laboratory (EMFL).

\appendix

\section{Samples and experimental setups}
\label{Sec:0}

\begin{figure*}[t!]
		\centering
		\includegraphics[width=0.8\linewidth]{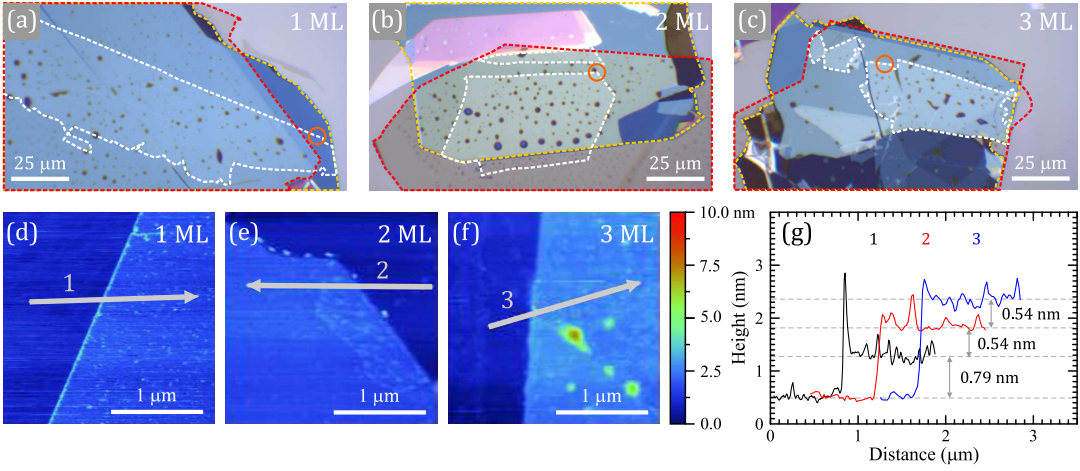}%
		\caption{Upper row: optical microscope images of hBN/MoS$_2$/hBN heterostructures comprising
         (a) mono-, (b) bi- and (c) trilayer MoS$_2$ flakes (outlined in white) sandwiched between two hBN flakes -- a 60-70 nm
         thick flake from the bottom (outlined in yellow) and 10-15 nm thick flake from the top (outlined in red). The spots of
         brown-to-blue colour visible in all images are bubbles of air and/or water vapour trapped between either the MoS$_2$ and
         the top hBN flake or the two hBN flakes. Lower row: (d)-(f) false colour atomic force microscope images taken at respective
         locations marked in (a)-(c) with orange circles. The grey arrows represent the line profiles drawn in panel (g) and labelled
         with the same numbers as in images (d)-(f).}
		\label{fig:fig_afm_results}
	\end{figure*}

In order to provide even and charged-defect-free support to van der Waals heterostructures studied in this work we assembled them on
ultraflat unoxidized silicon (Si) substrates from Ted Pella Inc., whose surface roughness amounts to about one-tenth of that of standard
Si/SiO$_2$ wafers. Each heterostructure consisted of three flakes obtained by means of mechanical exfoliation and successively stacked
one on top of another using dry transfer techniques: a 60-70 nm thick bottom hBN flake, a central MoS$_2$ flake of 1ML, 2ML or 3ML
thickness, and a 10-15 nm thick top hBN flake.

The optimum thicknesses of hBN flakes were estimated on the basis of transfer-matrix
simulations of reflectance contrast spectra of complete heterostructures, aiming at maximizing the visibility of absorption resonances
in MoS$_2$ flakes while keeping the thickness of bottom hBN flakes below 100 nm (a value up to which our exfoliation method yields a rich
harvest of flat, defect-free flakes with large-area terraces of constant thickness).

Using high-purity ELP BT-150E-CM tape from Nitto Denko,
bottom hBN flakes were exfoliated from high-quality bulk crystals grown under high-pressure conditions \cite{taniguchi2007} and then
transferred in a non-deterministic way onto clean and freshly annealed at $200^\circ\mathrm{C}$ Si substrates. The MoS$_2$ flakes were
obtained by means of two-stage, tape- and polydimethylsiloxane (PDMS)-based exfoliation of a bulk crystal purchased from HQ Graphene and
then stacked on the bottom hBN flakes using an all-dry deterministic stamping technique \cite{gomez}. The same procedure was applied also
to the top hBN flakes. While performing deterministic transfers, special attention was paid to mutual angular alignment of the flakes and
to immediate 10-minute-long after-transfer annealing of the samples at $180^\circ\mathrm{C}$ on a hot plate kept in clean ambient atmosphere.

Optical microscope images of investigated heterostructures which were fabricated in this way are shown in Fig.~\ref{fig:fig_afm_results} (a)-(c).
The spots of brown-to-blue colour visible in all images are up to 60-nm-high bubbles of air and/or water vapour (possibly with some amount of
hydrocarbons) trapped between either the MoS$_2$ and the top hBN flake or the two hBN flakes. Their appearance directly results from the
corrugation of thin, soft and flexible top hBN flakes supported by a PDMS stamp. Importantly, the areas between the bubbles, whose size
reaches up to 5 by 10 micrometers, exhibit very flat and high-quality intefraces between the constitutent flakes, as revealed with optical
measurements and atomic force microscopy (AFM) characterization.

An example of tapping-mode AFM imaging performed on finished heterostructures
with the use of Digital Instrument’s Dimension 3100 microscope is shown in Fig.~\ref{fig:fig_afm_results} (d)-(f). The false colour AFM images
correspond to respective locations marked in panels (a)-(c) with orange circles. The grey arrows represent the line profiles drawn in
Fig.~\ref{fig:fig_afm_results} (g) and labelled with the same numbers as in images (d)-(f). As can be seen, the profiles unambiguously
confirm the mono-, bi-, and trilayer thickness of MoS$_2$ flakes incorporated into the heterostructures shown in the upper row of
Fig.~\ref{fig:fig_afm_results}.

Compared to based on neutron scattering measurements estimation equal to 0.615 nm \cite{wakabayashi1975}, the
value of 0.79 nm we got for the monolayer most probably indicates that the equilibrium distance between the hBN and MoS$_2$ layers differs from
that between two neighboring MoS$_2$ layers in a bulk crystal. A smaller single-layer step hight obtained for the second and third layer in the
bi- and trilayer MoS$_2$ flakes may on the other hand imply the existence of small uniaxial compressive strain along the $c$-axis in hBN/MoS$_2$/hBN
heterostructures.

Measurements at zero magnetic field were carried out with the aid of a continuous flow cryostat
mounted on $x-y$ motorized positioners. The samples were placed on a cold finger of the cryostat. The temperature of the samples was kept
at $T=5$~K. The excitation light was focused by means of a 50x long-working distance objective with a 0.5 numerical aperture producing a
spot of about 1~$\mu$m.  The signal was collected via the same microscope objective, sent through a 0.5 m monochromator, and then detected
by a CCD camera.

Magneto-optical experiments were performed in the Faraday configuration using an
optical-fiber-based insert placed in a resistive or a superconducting magnet producing magnetic fields up
to 29~T or 14~T, respectively. The sample was mounted on top of an $x-y-z$ piezo-stage kept in gaseous
helium at $T$=4.2~K. The $\mu$-RC experiments were performed with the use of 100~W
tungsten halogen lamp. The excitation light was coupled to an optical fiber with a
core of 50 $\mu$m diameter and focused on the sample by an aspheric lens (spot
diameter around 10 $\mu$m). The signal was collected by the same lens, injected
into a second optical fiber of the same diameter, and analyzed by a 0.5~m long
monochromator equipped with a CCD camera. A combination of a quarter wave plate
and a polarizer are used to analyse the circular polarization of signals.

\section{Bilayer in magnetic field}
\label{Sec:A}

The magnetic field correction to the valence band Hamiltonian at the $\mathrm{K}^+$ point written in the basis $\{|\Psi^{(1)}_v\rangle \otimes|s\rangle,|\Psi^{(2)}_v\rangle\otimes|s\rangle\}$ is
$\delta H^{(2)}_{vs}=\big(G^{(2)}_v+\sigma_s\mathbb{1}\big)\mu_B B$. Here $\mu_B$ is the Bohr magneton, $B$ is the  magnetic
field, applied perpendicularly to the bilayer plane, $\mathbb{1}$ is the unit matrix and
\begin{equation}
G^{(2)}_v=
        \left[
        \begin{array}{cc}
            g_v-\delta g_v & 0 \\
            0 & -g_v+\delta g_v \\
          \end{array}
        \right].
\end{equation}
The correction to the conduction band Hamiltonian written in the basis
$\{|\Psi^{(1)}_c\rangle \otimes|s\rangle,|\Psi^{(2)}_c\rangle\otimes|s\rangle\}$ has the same form
$\delta H^{(2)}_c=\big(G^{(2)}_c+\sigma_s\mathbb{1}\big)\mu_B B$, where
\begin{equation}
G^{(2)}_c=
        \left[
        \begin{array}{cc}
            g_c-\delta g_c  & 0  \\
            0 & -g_c+\delta g_c  \\
          \end{array}
        \right]\mu_BB.
\end{equation}
The expressions for $g_v$, $g_c$, $\delta g_v$, $\delta g_c$ are derived in \cite{Arora2018}.
The corrections to the Hamiltonians provide the energy shifts of excitons in magnetic field.
The intralayer $X_A$ and $X_B$ excitons, constructed from the quasiparticles at the $\mathrm{K}^+$ point of bilayer,
are active in $\sigma^\pm$ polarizations. The corresponding energy shifts are linear in magnetic field
$\delta E_{A,B}^{\sigma^\pm}=\pm g^{(2)}_{A,B}\mu_B B$.
The $g$-coefficients of such exciton transitions are
\begin{align}
\label{eqs:A}
g_{A}^{(2)}=-g_u+(g_c-\delta g_c)-(g_v-\delta g_v)\frac{\Delta_v}{\sqrt{\Delta_v^2+4t^2}}, \\
\label{eqs:B}
g_{B}^{(2)}=g_u+(g_c-\delta g_c)-(g_v-\delta g_v)\frac{\Delta_v}{\sqrt{\Delta_v^2+4t^2}}.
\end{align}
We introduced $g_u=2m_0u^2/\hbar^2\Delta_c$ parameter, which originates from $k$-dependent admixture of
conduction bands. The mixing is negligibly small in the absence of magnetic field, but
gives finite correction if $B\neq0$.
The interlayer excitons $X_A'$ and $X_B'$ at the bilayer $\mathrm{K}^+$ point
are also active in $\sigma^\pm$ polarizations with  the magnetic shifts
$\delta E_{A',B'}^{\sigma^\pm}=\pm g^{(2)}_{A',B'}\mu_B B$ respectively.
The corresponding $g$-coefficients are
\begin{align}
\label{eqs:A'}
g_{A'}^{(2)}=g_u+(g_c-\delta g_c)+(g_v-\delta g_v)\frac{\Delta_v}{\sqrt{\Delta_v^2+4t^2}},\\
\label{eqs:B'}
g_{B'}^{(2)}=-g_u+(g_c-\delta g_c)+(g_v-\delta g_v)\frac{\Delta_v}{\sqrt{\Delta_v^2+4t^2}}.
\end{align}
Note that there are the following relations between $g$-coefficients of interlayer and intralayer exciton transitions
\begin{align}
&g^{(2)}_A+g^{(2)}_{A'}=g^{(2)}_B+g^{(2)}_{B'}=2(g_c-\delta g_c), \\
&g^{(2)}_{A'}-g^{(2)}_B=g^{(2)}_{B'}-g^{(2)}_A= (g_v-\delta g_v)\frac{2\Delta_v}{\sqrt{\Delta_v^2+4t^2}}, \\ &g^{(2)}_B-g^{(2)}_A=g^{(2)}_{A'}-g^{(2)}_{B'}=2g_u.
\end{align}
One can mention that the corresponding transitions at the $\mathrm{K}^-$ point in $\sigma^\pm$ polarizations
are characterized by the same values $\pm g^{(2)}_{A,B}$ and $\pm g^{(2)}_{A',B'}$ as in $\mathrm{K}^+$ point.
This is the consequences of time reversal and inversion symmetries of the crystal.
As a result, we can restore the absorption spectra of the bilayer in
$\sigma^-$ and $\sigma^+$ polarizations using our methodology of the experiment --- by  measuring the reflected light
in fixed $\sigma^+$ polarization and changing the orientation of magnetic field from $B$ to $-B$. The change of the
direction of the magnetic field to $-B$ in experiment mimics the transitions in $\sigma^-$ polarization of the reflected light.

We use the formula $g=[E^{\sigma^+}(B)-E^{\sigma^+}(-B)]/\mu_BB=[\delta E^{\sigma^+}(B)-\delta E^{\sigma^+}(-B)]/\mu_BB$
for exciton's $g$-factor. As a result the intralayer and interlayer $A,B$ excitons have $2g^{(2)}_{A,B}$ and $2g^{(2)}_{A',B'}$
$g$-factors, respectively.

The possible signs of intra- and interlayer exciton $g$-factors can be obtain from the expressions~(\ref{eqs:A}), (\ref{eqs:B}),
(\ref{eqs:A'}) and (\ref{eqs:B'}). Indeed, in the experiment we have $2g^{(2)}_{A,B}\approx -4$. It means that corrections $g_u$,
$\delta g_c$ and $\delta g_v$ are not significally large, and therefore very roughly $2g^{(2)}_{A,B}\approx 2(g_c -g_v)$ is nothing
but the $g$-factors of monolayer $A$ and $B$ excitons, for which we know that $g_v>g_c>0$. Substituting the positive $g_c$ and $g_v$
into Eqs.~(\ref{eqs:A'}) and (\ref{eqs:B'}) one can see that $2g^{(2)}_{A',B'}>0$, which is confirmed from the experiment.

\section{Trilayer in magnetic field}
\label{Sec:B}
The magnetic field correction to the valence and conduction band Hamiltonians written in the basis
$\{|\Psi^{(1)}_n\rangle\otimes|s\rangle,|\Psi^{(2)}_n\rangle\otimes|s\rangle,|\Psi^{(3)}_n\rangle\otimes|s\rangle\}$, is $\delta H^{(3)}_{ns}=\big(G^{(3)}_n+\sigma_s\mathbb{1}\big)\mu_B B$ where
\begin{equation}
G^{(3)}_n=
        \left[
        \begin{array}{ccc}
            g_n-\delta g_n & 0  & \bar{g}_n\\
            0 & -g_n+2\delta g_n & 0 \\
            \bar{g}_n & 0 & g_n-\delta g_n \\
          \end{array}
        \right],
\end{equation}
Here $n=v,c$ and $\bar{g}_c$ and $\bar{g}_v$ are the additional parameters which describe the magnetic dependent
coupling between layers of the system \cite{Arora2018}.
In new basis, defined in the main part of the text, the full Hamiltonians $H^{(3)}_{ns}+\delta H^{(3)}_{ns}$
are reduced to a block-diagonal form. Namely, the $G_n^{(3)}$ matrix transforms to
\begin{equation}
\mathcal{G}^{(3)}_n=
        \left[
        \begin{array}{ccc}
            g_n-\delta g_n+\bar{g}_n  & 0 & 0 \\
            0 & -g_n+2\delta g_n & 0 \\
            0 & 0 & g_n-\delta g_n-\bar{g}_n   \\
          \end{array}
        \right].
\end{equation}

The $1\times 1$ block corresponds to odd states $|\Upsilon^{(3)}_{ns}\rangle\otimes|s\rangle$
with total energies
\begin{equation}
E_{ns}^o(B)=E_n+\sigma_s\frac{\Delta_n}{2}+(g_n-\delta g_n-\bar{g}_n+\sigma_s)\mu_BB,
\end{equation}
with $E_v=0$ and $E_c=E_g$.
The expressions coincide with the monolayer ones, but with the new $g$-coefficient $g_n-\delta g_n+\bar{g}_n$.
Moreover, the $uk_\pm$ terms do not affect the odd states, and therefore such excitons have the same reduced
masses as their monolayer analogs. Hence the corresponding trilayer exciton line has the same optoelectronic
properties as it's monolayer analog. The odd $A$ and $B$ exciton transitions at the $\mathrm{K}^+$ point are
active only in $\sigma^+$ polarization. The corresponding energy shift in magnetic field is
$\delta E^{\sigma^+}_{A^o,B^o}=[(g_c-\delta g_c)-(g_v-\delta g_v)-\bar{g}_c+\bar{g}_v]\mu_BB$. The same type of
transitions at the $\mathrm{K}^-$ point are active in $\sigma^-$ polarization and have the energy shift
in magnetic field  $\delta E^{\sigma^-}_{A^o,B^o}=-\delta E^{\sigma^+}_{A^o,B^o}$. It immediately gives us
$g_{X^o_{A,B}}=2(g_c-g_v)-4\mathfrak{g}$ for $X^o_{A}$ and $X^o_{B}$ exciton $g$-factors both for $\mathrm{K}^+$
and $\mathrm{K}^-$ points of trilayer. Here we introduce the parameter
\begin{equation}
\mathfrak{g}=\frac12\Big(\delta g_c+\bar{g}_c-\delta g_v -\bar{g}_v\Big).
\end{equation}
From the experiment we know that $g_{X^o_{A}}\approx -4.5$, which is close to monolayer
$g_{X_A}=2(g_c-g_v)\approx -4$. Therefore, we can roughly estimate $\mathfrak{g}\approx 0.125$.

The second block is $2\times 2$ matrix, written in the basis of even states
$\{|\Upsilon^{(1)}_{n}\rangle\otimes|s\rangle, |\Upsilon^{(2)}_{n}\rangle\otimes|s\rangle\}$.
Note that the structure of this matrix does not coincide with the structure of $G_n^{(2)}$.
Therefore the subsystem of even states of trilayer demonstrates another behavior in magnetic field
than it's bilayer analog.
The conduction band states in $\text{K}^+$ point are decoupled and have the energies
\begin{align}
&E^{(1)}_{cs}=E_g+\sigma_s\frac{\Delta_c}{2}+(g_c-\delta g_c+\bar{g}_c+\sigma_s-2\sigma_sg_u)\mu_BB, \\
&E^{(2)}_{cs}=E_g-\sigma_s\frac{\Delta_c}{2} - (g_c-2\delta g_c-\sigma_s-2\sigma_sg_u)\mu_BB.
\end{align}
The conduction band energies correspond to the states
$\{|\Upsilon^{(1)}_{c}\rangle\otimes|s\rangle, |\Upsilon^{(2)}_{c}\rangle\otimes|s\rangle\}$. The valence band energies
of admixed states as a function of magnetic field $ B$ are calculated similarly to the bilayer case.

The energy shifts of intralayer $A$ and $B$ excitons in $\sigma^\pm$ polarizations at the $\mathrm{K}^+$ point
are $\delta E_{A,B}^{\sigma^\pm}=[\pm g^{(3)}_{A,B} + \mathfrak{g}]\mu_BB $, where
\begin{align}
g_{A}^{(3)}&=-2g_u+\Big(g_c-\frac32\delta g_c+\frac12\bar{g}_c\Big)- \nonumber \\
&-\Big(g_v-\frac32\delta g_v+\frac12\bar{g}_v\Big)\frac{\Delta_v}{\sqrt{\Delta_v^2+8t^2}}, \\
g_{B}^{(3)}&=2g_u+\Big(g_c-\frac32\delta g_c+\frac12\bar{g}_c\Big)- \nonumber \\
&-\Big(g_v-\frac32\delta g_v+\frac12\bar{g}_v\Big)\frac{\Delta_v}{\sqrt{\Delta_v^2+8t^2}}.
\end{align}
The energy shifts for interlayer $A'$ and $B'$ excitons in both polarizations have the form
$\delta E_{A',B'}^{\sigma^\pm}=[\pm g^{(3)}_{A',B'} + \mathfrak{g}]\mu_BB$, where
\begin{align}
g_{A'}^{(3)}&=2g_u+\Big(g_c-\frac32\delta g_c+\frac12\bar{g}_c\Big)+ \nonumber \\
&+\Big(g_v-\frac32\delta g_v+\frac12\bar{g}_v\Big)\frac{\Delta_v}{\sqrt{\Delta_v^2+8t^2}}, \\
g_{B'}^{(3)}&=-2g_u+\Big(g_c-\frac32\delta g_c+\frac12\bar{g}_c\Big)+ \nonumber \\
&+\Big(g_v-\frac32\delta g_v+\frac12\bar{g}_v\Big)\frac{\Delta_v}{\sqrt{\Delta_v^2+8t^2}}.
\end{align}
In the absence of $\mathfrak{g}$ the latter results coincide with the bilayer case up to
redefinition of the parameters. The non-zero value of $\mathfrak{g}$ shows remarkable difference
between pure bilayer and effective bilayer cases. The corresponding energy shifts in $\sigma^\pm$
polarizations at the $\mathrm{K}^-$ point are $\delta E_{A,B}^{\sigma^\pm}=[\pm g_{A,B}^{(3)} - \mathfrak{g}]\mu_BB$
and $\delta E_{A',B'}^{\sigma^\pm}=[\pm g^{(3)}_{A',B'} - \mathfrak{g}]\mu_BB$.

This non-equivalency makes the analysis of the $g$-factors of the system more complicated. Let us consider the
results of the measurements presented on the Fig.~\ref{fig:fig_2} more carefully, focusing mainly on $X_A$ and
$X'_A$ exciton transitions. Again, according to the methodology of our experiment we measure the $g$-factor using
the formula $g=[\delta E^{\sigma^+}(B)-\delta E^{\sigma^+}(-B)]/\mu_BB$. Then the $g$-factors at the $\mathrm{K}^+$
point of intralayer $X_{A}$, $X_{B}$ and interlayer $X'_{A}$, $X'_{B}$ excitons are
$g^{\mathrm{K}^+}_{X_{A,B}}=2g^{(3)}_{A,B}+2\mathfrak{g}$ and $g^{\mathrm{K}^+}_{X_{A',B'}}=2g^{(3)}_{A',B'}+2\mathfrak{g}$
respectively. For $\mathrm{K}^-$ point transitions we obtain $g^{\mathrm{K}^-}_{X_{A,B}}=2g^{(3)}_{A,B}-2\mathfrak{g}$ and $g^{\mathrm{K}^-}_{X_{A',B'}}=2g^{(3)}_{A',B'}-2\mathfrak{g}$. Taking into account the relative smallness of $\mathfrak{g}$
and absence of the results for magnetic field $B$ larger than $14$~T we suppose that the double $g$-factor structure of $X_A$
and $X'_A$ resonances is indistinguishable. Instead of this, probably, we observe only their average values
$g_{X_{A,B}}=2g^{(3)}_{A,B}$ and $g_{X_{A',B'}}=2g^{(3)}_{A',B'}$ respectively. One can mention that these average $g$-factors
surprisingly coincide with the ones we can get from the standard formula
$g=[E^{\sigma^+}(B)-E^{\sigma^-}(B)]/\mu_BB=[\delta E^{\sigma^+}(B)-\delta E^{\sigma^-}(B)]/\mu_BB$. The analysis of the signs of
intra- and interlayer exciton $g$-factors can be done in the same way as in bilayer case.

\end{document}